\documentstyle[preprint,eqsecnum,aps,pra]{revtex}
\begin{document}
\preprint{DAMTP-96-89, quant-ph/9611054}

\title{Maximum Information and Quantum Prediction Algorithms}

\author{Jim McElwaine \thanks{E-mail:jnm11@damtp.cam.ac.uk}}

\address{Department of Applied Mathematics and Theoretical Physics,\\
  University of Cambridge,\\ Silver Street, Cambridge CB3 9EW, U.K.}

\date{28th November, 1996} \maketitle
\begin{abstract}
  This paper describes an algorithm for selecting a consistent set
  within the consistent histories approach to quantum mechanics and
  investigates its properties. The algorithm uses a maximum
  information principle to select from among the consistent sets
  formed by projections defined by the Schmidt decomposition. The
  algorithm unconditionally predicts the possible events in closed
  quantum systems and ascribes probabilities to these events.  A
  simple spin model is described and a complete classification of all
  exactly consistent sets of histories formed from Schmidt projections
  in the model is proved. This result is used to show that for this
  example the algorithm selects a physically realistic set.  Other
  tentative suggestions in the literature for set selection algorithms
  using ideas from information theory are discussed.
\end{abstract}

\vspace{15pt}
\pacs{PACS numbers: 03.65.Bz, 98.80.H \\[10pt] Submitted to Phys. Rev. A}

\section{Introduction}\label{sec:inf:intro} 

It is hard to find an entirely satisfactory interpretation of the
quantum theory of closed systems, since quantum theory does not
distinguish physically interesting time-ordered sequences of
operators. In this paper, we consider one particular line of attack on
this problem: the attempt to select consistent sets by using the
Schmidt decomposition together with criteria intrinsic to the
consistent histories formalism. For a discussion of why we believe
consistent histories to be incomplete without a set selection
algorithm see~\cite{Dowker:Kent:properties,Dowker:Kent:approach} and
for other ideas for set selection algorithms
see~\cite{McElwaine:Kent,McElwaine:2,%
gmhstrong,Isham:Linden:information}.
This issue is controversial: others believe that the consistent
histories approach is complete in
itself~\cite{omnesbook,griffithschqr,CEPI:GMH}.

\subsection{Consistent histories formalism}\label{ssec:CHformalism} 

We use a version of the consistent histories formalism in which the
initial conditions are defined by a pure state, the histories are
branch-dependent and consistency is defined by Gell-Mann and Hartle's
medium consistency criterion eq.~(\ref{mediumcon}).  We restrict
ourselves to closed quantum systems with a Hilbert space in which we
fix a split ${\cal H} = {\cal H}_1 \otimes {\cal H}_2$; we write $\dim
({\cal H}_j ) = d_j$ and we suppose that $d_1 \leq d_2 < \infty$.  The
model described in sec.~\ref{sec:spin:intro} has a natural choice for
the split. Other possibilities are discussed in~\cite{McElwaine:Kent}.

Let $|\psi\rangle$ be the initial state of a quantum system.  A
\emph{branch-dependent set of histories} is a set of products of
projection operators indexed by the variables $\alpha = \{ \alpha_n ,
\alpha_{n-1} , \ldots , \alpha_1 \}$ and corresponding time
coordinates $\{ t_n , \ldots , t_1 \}$, where the ranges of the
$\alpha_k$ and the projections they define depend on the values of
$\alpha_{k-1} , \ldots , \alpha_1 $, and the histories take the form:
\begin{equation} \label{histories}
  C_{\alpha} = P_{\alpha_n}^n (t_n ; \alpha_{n-1} , \ldots , \alpha_1)
  P_{\alpha_{n-1}}^{n-1} (t_{n-1} ; \alpha_{n-2} , \ldots , \alpha_1)
  \ldots P_{\alpha_1}^1 ( t_1 )\,.
\end{equation}
Here, for fixed values of $\alpha_{k-1} , \ldots , \alpha_1$, the
$P^k_{\alpha_k} (t_k ; \alpha_{k-1} , \ldots , \alpha_1 )$ define a
projective decomposition of the identity indexed by $\alpha_k$, so
that $\sum_{\alpha_k} P^k_{\alpha_k} (t_k ; \alpha_{k-1} , \ldots ,
\alpha_1 ) = 1 $ and
\begin{equation} \label{decomp}
P^k_{\alpha_k} (t_k ; \alpha_{k-1} , \ldots , \alpha_1 )
P^k_{\alpha'_k} (t_k ; \alpha_{k-1} , \ldots , \alpha_1 ) =
\delta_{\alpha_k \alpha'_k } P^k_{\alpha_k} (t_k ; \alpha_{k-1} ,
\ldots , \alpha_1 )\,.
\end{equation}
Here and later, though we use the compact notation $\alpha$ to refer
to a history, we intend the individual projection operators and their
associated times to define the history.

We use the consistency criterion\footnote{For a discussion of other
  consistency criteria see, for example,
  refs.~\cite{Kent:implications,Goldstein:Page,%
    Dowker:Halliwell,McElwaine:1}.}
\begin{equation}
  D_{\alpha\beta} = 0, \quad \forall \alpha \neq \beta,
  \label{mediumcon}
\end{equation}
which Gell-Mann and Hartle call \emph{medium consistency}, where
$D_{\alpha\beta}$ is the \emph{decoherence matrix}
\begin{equation}
  D_{\alpha\beta} = \mbox{Tr}\, (C_\alpha \rho C_\beta^\dagger)\,.
  \label{dmdef}
\end{equation}
Probabilities for consistent histories are defined by the formula
\begin{equation}\label{probdef}
  p(\alpha) = D_{\alpha\alpha}.
\end{equation}

With respect to the ${\cal H} = {\cal H}_1 \otimes {\cal H}_2$
splitting of the Hilbert space, the \emph{Schmidt decomposition} of
$|\psi (t) \rangle$ is an expression of the form
\begin{equation} \label{schmidteqn}
  |\psi (t) \rangle = \sum_{i=1}^{d_1} \, [p_i(t)]^{1/2} \, | w_i
  (t)\rangle_1 \otimes |w_i (t)\rangle_2 \, ,
\end{equation}
where the \emph{Schmidt states} $\{ |w_i\rangle_1 \}$ and $\{
|w_i\rangle_2\}$ form, respectively, an orthonormal basis of ${\cal
H}_1$ and part of an orthonormal basis of ${\cal H}_2$, the functions
$p_i (t)$ are real and positive, and we take the positive square root.
For fixed time $t$, any decomposition of the form
eq.~(\ref{schmidteqn}) then has the same list of probability weights
$\{ p_i (t) \}$, and the decomposition~(\ref{schmidteqn}) is unique if
these weights are all different. These probability weights are the
eigenvalues of the reduced density matrix.

The idea motivating this paper is that the combination of the ideas of
the consistent histories formalism and the Schmidt decomposition might
allow us to define a mathematically precise and physically interesting
description of the quantum theory of a closed system. We consider
constructing histories from the projection operators\footnote{There
are other ways of constructing projections from the Schmidt
decomposition~\cite{McElwaine:Kent}, though for the model considered
in this paper the choices are equivalent.}
\begin{equation} \label{schmidtprojs1}
  \begin{array}{lll}
  P_i (t) = | w_i (t) \rangle_1 \langle w_i (t) |_1 \otimes I_2
  &\mbox{and}& \overline P = I_1 \otimes I_2 - \sum_i P_i (t)\, ,
 \end{array}
\end{equation}
which we refer to as \emph{Schmidt projections}. If $\mbox{dim}{\cal
  H}_1 = \mbox{dim}{\cal H}_2$ the complementary projection $\overline
  P$ is zero. In developing the ideas of this paper, we were
  influenced in particular by Albrecht's
  investigations~\cite{Albrecht:decoherence,Albrecht:collapsing} of
  the behaviour of the Schmidt decomposition in random Hamiltonian
  interaction models and the description of these models by consistent
  histories.

\section{Information}
Recent work~\cite{McElwaine:Kent,Isham:Linden:information,%
  McElwaine:4,Hartle:spacetime:information} has shown some of the
difficulties in formulating a successful set selection algorithm.  The
analysis of~\cite{McElwaine:Kent} suggests that in many systems no
algorithm that constructs sets by proceeding forwards in time will
produce the correct physical set. If so, an algorithm must consider
the entire time evolution of a system if it is always to overcome this
problem. This paper introduces an algorithm that is global with
respect to time: the algorithm considers the
class\footnote{\emph{Class} is used as a synonym for \emph{set} when
referring to a \emph{set} of sets of consistent histories.} of all
consistent sets of histories formed from Schmidt projections and
selects from among them the one with the greatest Shannon
information~\cite{Shannon:Weaver}.

Information\footnote{\emph{Entropy} or \emph{information-entropy} are
  used instead by some authors.} is a term often used in the study of
  quantum mechanics and is used in many different senses.
  Hartle~\cite{Hartle:spacetime:information} \footnote{For comments on
  and corrections to Hartle`s paper see \cite{Kent:information}}
  considers the \emph{missing information} of a set of histories in a
  generalised spacetime quantum mechanics --- he defines the missing
  information S of a set of histories ${\cal S}$ with initial density
  matrix $\rho$ as
\begin{equation}\label{inf:deficit}
  S({\cal S},\rho) = \max_{\rho'\in\{D({\cal S},\rho') = D( {\cal
    S},\rho)\}} E(\rho')\,,
\end{equation}
where $D({\cal S},\rho)$ is the decoherence matrix for the set of
histories ${\cal S}$ with initial density matrix $\rho$. Throughout
this paper $E$ will denote the Shannon information of a set of
probabilities or, in the case of a positive definite Hermitian matrix,
the Shannon information of its eigenvalues\footnote{in information
theory the singularity for zero probabilities is removed by defining
$0 \log 0 = 0$.}. So, for example, $E(\rho') = - \mbox{Tr} \rho' \log
\rho'$ and
\begin{equation}
  E({\cal S},\rho) = \sum_{\alpha \in {\cal S}} - D_{\alpha\alpha}
  \log D_{\alpha\alpha}\,,
\end{equation}
where $\{D_{\alpha\alpha}\}$ are the diagonal elements of the
decoherence matrix $D({\cal S},\rho)$.  Note that if a set of
histories ${\cal S}$ is medium consistent then $E({\mathcal S},\rho) =
E[D({\cal S},\rho)]$: generically this is not true for weak
consistency criteria.

$S({\cal S},\rho)$ is the information content of a
\emph{maximum-entropy}~\cite{Jaynes:papers} estimation of the initial
density matrix given the set of histories and their probabilities---
it quantifies what can be inferred about the initial density matrix
using the set of histories and their probabilities.  Hartle goes on to
define
\begin{equation}\label{inf:min}
  S({\cal G},\rho) = \min_{{\cal S}\in {\cal G}} S({\cal S},\rho),
\end{equation}
where ${\cal G}$ is some class of consistent sets of histories.
Computing $S({\cal G},\rho)$ for different classes enables one to
understand different ways information about a quantum system can be
obtained. For example Hartle suggests comparing whether the same
information is available using homogeneous~\cite{Isham:homog}
histories instead of the more general inhomogeneous histories. When
${\cal G}$ is the class of all consistent sets he calls $S({\cal
G},\rho)$ the \emph{complete information}.

Eq.~(\ref{inf:min}) could be used as the basis for a set selection
algorithm by specifying some class of sets of histories ${\cal G}$ and
selecting a set of histories that produces the minimum in
eq.~(\ref{inf:min}). This does not work for general classes, since if
the class contains sets of histories which include projections onto
the eigenspaces of $\rho$ (in non-relativistic quantum mechanics)
these projections completely specify $\rho$, so a rather uninteresting
set of histories is selected. However, if the initial state is pure
and a \emph{Schmidt class} (a class of sets of histories formed from
Schmidt projections) is used it will not generically contain a set of
histories that includes a rank one projection onto the initial state,
hence the set of histories selected by eq.~(\ref{inf:min}) might not
be trivial.  For instance the set of histories consisting of
projections $P \otimes I$ and $\overline P \otimes I$, where $P$ is
the projection onto the non-zero system Schmidt eigenspaces, has
missing information $\log \mbox{rank}(P \otimes I)$. It might be
considered unnatural to assume a pure initial state and then make a
maximum entropy calculation over density matrices of other ranks;
however, this idea has a more serious flaw. The aim of set selection
algorithms is to make statements concerning physical events, not
merely to supply initial conditions. This algorithm only searches for
a set of histories that best specifies the initial conditions and
there is no reason to expect it to produce sets that do more than
describe the initial conditions.

Isham and Linden~\cite{Isham:Linden:information} independently,
recently proposed a different version of missing information, which
they call \emph{information-entropy}, that is simpler and does not use
ideas of maximum entropy.
\begin{equation}\label{IL:information}
  S'({\cal S},\rho) = - \sum_{\alpha \in {\cal S}} D_{\alpha\alpha}
  \log \frac{D_{\alpha\alpha}}{{\widehat{\mbox{dim}}}^2\ (\alpha)}\,,
\end{equation}
where
\begin{equation}
  \widehat{\mbox{dim}}\ \alpha =
  \frac{\mbox{Tr}(C_\alpha)}{\mbox{Tr}(I)}
\end{equation}
is the \emph{normalised dimension} of the history, and $C_\alpha$ and
$I$ are considered as operators in the same $n$-fold tensor product
space~\cite{Isham:Linden:temporal,Isham:Linden:Schreckenberg} of
${\cal H}$.  For example, if the history $\alpha$ is defined by
consecutive projections $\{P_k, k = 1,\ldots,n\}$ then
$\widehat{\mbox{dim}}\ (\alpha) = \mbox{Tr} (P_1 \otimes \cdots
\otimes P_n) / d^n= \mbox{rank}(P_1) \times \cdots \times
\mbox{rank}(P_n)/d^n$.  Like Hartle's missing information, $S'$
\emph{decreases} under refinements and extensions of ${\cal S}$.
Isham and Linden show that
\begin{equation}\label{IL:lbound}
  \min_{{\cal S} \in {\cal G}} S' ({\cal S},\rho) \geq -\mbox{Tr $\rho
  \log \rho$}- n \log d
\end{equation}
and for some examples that the bound is obtained, and they conjecture
that the bound is attained in general. Isham and Linden also suggest
that information-entropy might help in the development of a set
selection criterion --- they suggest that perhaps the minimisation
should be carried out with respect to a system--environment split.
Clearly some restriction on the class of sets used is necessary since
bound~(\ref{IL:lbound}) contains no mention of the Hamiltonian or time
evolution of the system --- simply minimising information-entropy is
unlikely to produce a good set selection algorithm, since the sets of
histories that describes experimental situations are much more than a
description of the initial conditions.

Gell-Mann and Hartle discuss similar ideas in detail in
ref.~\cite{gmhstrong}. They introduce a measure, which they call
\emph{total information} or \emph{augmented entropy}, $\Sigma$ that
combines algorithmic information (see for example
ref.~\cite{Zurek:algorithmic}), entropy-information and coarse
graining. This is an attempt to provide a quantitative measure of
\emph{quasiclassicality}. They show that minimising $\Sigma$ does not
provide a useful set selection algorithm --- the results are trivial,
histories are selected that consist of nothing but projections onto
the initial state --- but they suggest augmenting the minimisation
with a stronger consistency criterion,
\begin{equation}\label{GMH:strong}
  \langle\alpha| M^\dagger_\alpha M_\beta |\beta\rangle = p_\alpha
  \delta_{\alpha\beta} \mbox{~$\forall \alpha \neq \beta$, $M_\alpha
  \in {\cal M}_\alpha$ and $M_\beta \in {\cal M}_\beta$,}
\end{equation}
where ${\cal M}_\alpha$ and ${\cal M}_\beta$ are sets of operators.
This is an interesting idea. So far however, Gell-Mann and Hartle have
not proposed a definite algorithm for choosing the ${\cal M}_\alpha$.
Without a concrete scheme for choosing the sets ${\cal M}_\alpha$ the
set selection problem of course becomes the problem of selecting
${\cal M}_\alpha$.  There seems a risk that Gell-Mann and Hartle's
proposal also has the previously mentioned disadvantage of favouring
set of histories that only provide a description of the initial state
and say nothing about the dynamics, though perhaps with a suitable
choice for ${\cal M}_\beta$ this problem would not arise.

The approach we present here starts with a precisely defined class of
quasiclassical sets of histories (formed from Schmidt projections) and
picks the set of histories from this class with the maximum
information.

It might seem counterintuitive to use a maximum information principle,
especially as other approaches in the literature to date have looked
at minimising measures of information.  However, these approaches have
started with a much larger class of sets of histories. Picking the set
with largest information from these classes would result in a
non-quasiclassical set of histories with each history having the same
probability. In this approach though, we are using a highly restricted
class --- the class formed using Schmidt projections. This class of
histories is so restricted that in some cases it may only consist of
sets with projections at $t=0$ onto the initial Schmidt states.
Picking the set with the largest information tends to pick the set
with the largest number of histories. Other functions of the history
probabilities could also be used, the essential requirement being that
the functions tend to increase with the number of projections. We
regard this proposal as a starting point for further investigations
into set selection algorithms --- especially since there are only
pragmatic rather than fundamental reasons for choosing maximum
information as a set selection axiom.

\section{Algorithm}\label{sec:inf:algorithm} 

Let ${\cal G}({\cal H},U,|\psi\rangle)$ be the class of all sets of
non-trivial\footnote{In this paper we call a history trivial if its
probability is zero and non-trivial if its probability is non-zero.},
exactly consistent, branch-dependent\footnote{A branch-independent
version of the algorithm can be formulated similarly} histories formed
{}from Schmidt projection operators, where ${\cal H} = {\cal H}_1
\otimes {\cal H}_2$ is a finite Hilbert space, $U(t)$ a time evolution
operator and $|\psi\rangle$ the initial state.  Note that in this
section the set of histories includes the initial state.  The
algorithm selects the set ${\cal S} \in {\cal G}$ with the greatest
Shannon information. That is
\begin{equation} \label{maxinf}
   \max_{{\cal S} \in {\cal G}} E({\cal S}) = \max_{{\cal S} \in {\cal
   G}} \sum_{\alpha \in {\cal S}} - p_\alpha \log p_\alpha,
\end{equation}
where $p_\alpha$ is the probability of history $\alpha$. The class
${\cal G}$ could be chosen differently by using any of the consistency
or non-triviality criteria from ref.~\cite{McElwaine:Kent}.  Another
variant uses sets of histories formed by Schmidt projections onto the
system eigenspaces of the individual path-projected-states
($U(t)C_\alpha|\psi\rangle$), not the total state, so that the choice
of projections is branch-dependent as well as the choice of projection
times. This is likely to be necessary in general to produce realistic
sets.

When the initial state is pure, in a Hilbert space of dimension $d$
($= d_1 d_2$) there can only be $d$ non-trivial, exactly consistent
histories within a set\footnote{There can be $2d$ if weak consistency
is used.}. In realistic examples approximate consistency may have to
be considered.  To ensure the algorithm is well defined it is
important that the number of possible histories is finite, which will
only be true if we use a parameterised non-triviality criterion or we
use a consistency criterion, such as the DHC, that can only be
satisfied by a finite number of histories~\cite{McElwaine:1}. This is
a natural requirement for any set of histories in a finite Hilbert
space since the exactly consistent sets are finite.

To show that the maximum in eq.~(\ref{maxinf}) exists we define two
sets of histories as \emph{information equivalent}, ${\cal S}_1 \sim
{\cal S}_2$, if $E({\cal S}_1) = E({\cal S}_2)$; that is, sets of
histories are information equivalent if they have the same
information.  Note that information equivalent sets generically are
not physically equivalent, but physically equivalent sets are
information equivalent.  Eq.~(\ref{maxinf}) selects an information
equivalent class of sets of histories that all have the maximum
information.  Sufficient conditions for eq.~(\ref{maxinf}) to be well
defined are that ${\cal G}/\kern-.35em\sim$ is closed and that
$E({\cal S})$ is bounded.  ${\cal G}$ itself is not closed, but the
only limit sets of histories it does not include are those containing
zero probability histories, and since zero probability histories
contribute zero information these limit sets are equivalent to sets
which are in ${\cal G}$, hence ${\cal G}/\kern-.35em\sim$ is closed.
Moreover these limit sets are also physically equivalent to some of
the sets that they are information equivalent to, since they only
differ by zero probability histories --- excluding the limit sets does
not change anything physical. The information of any set of histories
in ${\cal G}$ is bounded, since the number of histories in any set of
histories in ${\cal G}$ is bounded and the information of a set of $n$
probabilities is bounded by $\log n$.  Conditions sufficient to ensure
uniqueness are much more complicated.  It seems likely that a unique
physically equivalent class will generically be selected, but in
special cases it is clear that this is not the case.

First we describe some useful properties of this algorithm and then we
apply it to a simple model.

\subsection{Completeness}\label{sec:completeness}

The set of histories selected by the algorithm cannot be extended
(except trivially) because any non-trivial extension increases the
information content. To see this consider the set of histories
${\mathcal S}$ and an extension ${\cal S}'$. The probabilities for the
new histories can be written in the form $p_\alpha q^{(\alpha)}_\beta$
where $\sum_\beta q^{(\alpha)}_\beta=1$ for all $\alpha$. The
information of the new set is
\begin{equation}\label{infadd}
  E({\cal S}') = -\sum_\alpha \sum_\beta p_\alpha q^{(\alpha)}_\beta
  \log p_\alpha q^{(\alpha)}_\beta = E({\cal S}) + \sum_\alpha
  p_\alpha E(q^{(\alpha)}_\beta),
\end{equation}
which is strictly greater than $E({\cal S})$ whenever the extension
results in at least one non-zero probability.

\subsection{Additivity}

A set of branch-dependent histories has a probability tree structure,
where each history $\alpha$ refers to a terminal node of the tree and
the unique path from that node to the root node. The nodes themselves
are associated with projection operators and path projected states.
Define ${\cal S}_{\alpha k}$ to be the set of all histories extending
{}from the $k^{\mbox{\scriptsize th}}$ node of history $\alpha$,
normalised so that the total probability is one. This is a set of
histories in its own right which will be consistent if the entire set
of histories is consistent. Consider a simple example where the first
projection produces two histories with probabilities $p$ and $q$ and
the subtrees from these nodes are ${\mathcal S}_p$ and ${\cal
S}_q$. The information for the set of histories can then be written,
\begin{equation}\label{subtreeadd}
  E({\cal S}) = E(\{p,q\}) + p E({\cal S}_p) + q E({\cal S}_q).
\end{equation}
This formula is easy to generalise. Each subtree must have maximum
information subject to the constraint that the history vectors span a
space orthogonal to the other history states. That is, a global
maximum must also be a local maximum in each degree of freedom and the
subtrees are the degrees of freedom.

\subsection{Large sets}

One of the problems with the algorithms in ref.~\cite{McElwaine:Kent}
is their tendency to make projections too early so that they prevent
projections at later times. Other problems also arise with algorithms
that produce histories with zero or small probabilities. The
maximum-information algorithm will not have these problems, since any
projection that prevents later extensions is unlikely to be selected,
histories with zero probability will never be selected (since they
contribute no information), and histories with small probabilities are
also unlikely to be selected. Therefore the algorithm is likely to
produce large complicated sets of histories.

\subsection{Stability}

It is difficult to prove any general results about stability for this
algorithm, but it seems likely to produce stable predictions for the
following reason.  The Schmidt projections and hence decoherence
matrix elements generically will vary continuously with sufficiently
small changes in the initial state and Hamiltonian, thus the algorithm
can be regarded as a continuous optimisation problem, and the
solutions to continuous optimisation problems are stable.

\section{A simple spin model}\label{sec:spin:intro} 

We now consider a simple model in which a single spin half particle,
the system, moves past a line of spin half particles, the environment,
and interacts with each in turn.  This can be understood as modelling
either a series of measurement interactions in the laboratory or a
particle propagating through space and interacting with its
environment.  In the first case the environment spin half particles
represent pointers for a series of measuring devices, and in the
second they could represent, for example, incoming photons interacting
with the particle.

Either way, the model omits features that would generally be
important.  For example, the interactions describe idealised sharp
measurements --- at best a good approximation to real measurement
interactions, which are always imperfect.  The environment is
represented initially by the product of $N$ particle states, which are
initially unentangled either with the system or each other.  The only
interactions subsequently considered are between the system and the
environment particles, and these interactions each take place in
finite time.  We assume too that the interactions are distinct: the
$k^{\mbox{\scriptsize th}}$ is complete before the
$(k+1)^{\mbox{\scriptsize th}}$ begins.

\subsection{Definition of the model} 

We use a vector notation for the system states, so that if ${\bf u}$
is a unit vector in $R^3$ the eigenstates of $\sigma. {\bf u }$ are
represented by $| \bf \pm u \rangle$.  With the pointer state analogy
in mind, we use the basis $\{ |\uparrow\rangle_k ,
|\downarrow\rangle_k \}$ to represent the $k^{\mbox{\scriptsize th}}$
environment particle state, together with the linear combinations
$|\pm\rangle_k = (|\uparrow\rangle_k \pm i|\downarrow\rangle_k
)/\sqrt{2}$.  We compactify the notation by writing environment states
as single kets, so that for example $ |\uparrow\rangle_1 \otimes
\cdots \otimes |\uparrow\rangle_n $ is written as $| \uparrow_1 \ldots
\uparrow_n \rangle$, and we take the initial state $|\psi(0)\rangle$
to be $|{\bf v}\rangle \otimes | \uparrow_1 \ldots \uparrow_n
\rangle$.

The interaction between the system and the $k^{\mbox{\scriptsize th}}$
environment particle is chosen so that it corresponds to a measurement
of the system spin along the ${\bf u}_k$ direction, so that the states
evolve as follows:
\begin{eqnarray} \label{measurementinteraction}
  |{\bf u}_k\rangle \otimes |\uparrow\rangle_k & \to & |{\bf
    u}_k\rangle \otimes |\uparrow\rangle_k \, , \\ |{\bf -u}_k \rangle
    \otimes |\uparrow\rangle_k & \to & |{\bf -u}_k \rangle \otimes
    |\downarrow\rangle_k.
\end{eqnarray}
A simple unitary operator that generates this evolution is
\begin{equation}\label{Ukdef} 
  U_k( t ) = P({\bf u}_k) \otimes I_k + P({\bf -u}_k) \otimes
  \mbox{e}^{-i\theta_k(t) F_k} \, ,
\end{equation}
where $P({\bf x}) = |{\bf x}\rangle \langle{\bf x}|$ and $F_k =
i|\downarrow\rangle_k \langle\uparrow|_k - i |\uparrow\rangle_k
\langle\downarrow|_k$.  Here $\theta_k(t)$ is a function defined for
each particle $k$, which varies from $0$ to $\pi/2$ and represents how
far the interaction has progressed.  We define $P_k (\pm) = |\pm
\rangle_k \langle\pm|_k $, so that $F_k = P_k (+)-P_k (-)$.

The Hamiltonian for this interaction is thus
\begin{equation}
  H_{k}(t) = i\dot U_k (t) U_k^\dagger (t) \\ = \dot \theta_k(t)
  P({\bf -u}_k) \otimes F_k \, ,
\end{equation}
in both the Schr\"odinger and Heisenberg pictures.  We write the
extension of $U_k$ to the total Hilbert space as
\begin{equation}\label{Vkdef} 
  V_k = P({\bf u}_k) \otimes I_1 \otimes \cdots \otimes I_n + P({\bf
    -u}_k) \otimes I_1 \otimes \cdots \otimes I_{k-1} \otimes
    \mbox{e}^{-i\theta_k(t) F_k} \otimes I_{k+1} \otimes \cdots
    \otimes I_n \,.
\end{equation}
We take the system particle to interact initially with particle $1$
and then with consecutively numbered ones, and there is no interaction
between environment particles, so that the evolution operator for the
complete system is
\begin{equation}
  U(t) = V_n(t) \ldots V_1(t) \, ,
\end{equation}
with each factor affecting only the Hilbert spaces of the system and
one of the environment spins.

We suppose, finally, that the interactions take place in disjoint time
intervals and that the first interaction begins at $t=0$, so that the
total Hamiltonian is simply
\begin{equation} 
 H (t ) = \sum_{k=1}^n H_k (t) \, ,
\end{equation} 
and we have that $\theta_1 (t) > 0 $ for $t > 0$ and that, if
$\theta_k(t) \in ( 0,\pi/2 ) $, then $\theta_i(t) = \pi/2
{\rm~for~all~} i < k$ and $\theta_i(t) = 0 {\rm~for~all~} i >k$.

\section{Classification of Schmidt projection consistent 
  sets in the model} 
\label{sec:spin:analysis}

For generic choices of the spin measurement directions, in which no
adjacent pair of the vectors $\{{\bf v},{\bf u}_1, \ldots ,{\bf
u}_n\}$ is parallel or orthogonal, the exactly consistent
branch-dependent sets defined by the Schmidt projections onto the
system space can be completely classified in this model. The following
classification theorem is proved in this section:

\vspace{.5\baselineskip}
\noindent\emph{Theorem}\qquad 
In the spin model defined above, suppose that no adjacent pair of the
vectors $\{{\bf v},{\bf u}_1, \ldots ,{\bf u}_n\}$ is parallel or
orthogonal.  Then the histories of the branch-dependent consistent
sets defined by Schmidt projections take one of the following forms:
\begin{description} 
\item[(i)] a series of Schmidt projections made at times between the
interactions --- i.e.\ at times $t$ such that $\theta_k (t) = {0
{\rm~or~} \pi/2} {\rm~for~all~} k$.
\item[(ii)] a series as in (i), made at times $t_1 , \ldots , t_n$,
together with one Schmidt projection made at any time $t$ during the
interaction immediately preceding the last projection time $t_n$.
\item[(iii)] a series as in (i), together with one Schmidt projection
made at any time $t$ during an interaction taking place after $t_n$.
\end{description} 
Conversely, any branch-dependent set, each of whose histories takes
one of the forms (i)-(iii), is consistent.  \vspace{.5\baselineskip}

\noindent We assume below that the set of 
spin measurement directions satisfies the condition of the theorem:
since this can be ensured by an arbitrarily small perturbation, this
seems physically reasonable.  The next sections explain, with the aid
of this classification, the results of various set selection
algorithms applied to the model.

\subsection{Calculating the Schmidt states}

Eq.~(\ref{Ukdef}) can be written
\begin{equation}
  U_j(t) = e^{-i\theta_j(t) P(-{\bf u}_j)} \otimes P_j(+) +
    e^{i\theta_j(t) P_j(-{\bf u}_j)} \otimes P_j(-).
\end{equation}
Define $x_{+j}(t) = \exp[-i\theta_j(t) P({\bf -u}_j)]$ and $x_{-j}(t)
= x^\dagger_{+j}(t)$ so $U_j(t) = x_{+j}(t) \otimes P_j(+) + x_{-j}(t)
\otimes P_j(-)$. Let ${\bf \pi}$ be a string of $n$ pluses and
minuses, $|\pi\rangle$ denote the environment state $|\pi_1\rangle_1
\otimes \cdots \otimes |\pi_n\rangle_n$, $P(\pi) =
|\pi\rangle\langle\pi|$ and $x_\pi(t) = x_{\pi_nn}(t) \ldots
x_{\pi_11}(t)$.  Then
\begin{equation}\label{Uteq}
  U(t) = \sum_\pi x_\pi(t) \otimes P(\pi).
\end{equation}
The time evolution of the initial state $|\psi(0)\rangle = |{\bf
  v}\rangle \otimes |\uparrow_1 \ldots \uparrow_n\rangle$, the
  corresponding reduced density matrix and the Schmidt decomposition
  can now be calculated,
\begin{equation}
  |\psi(t)\rangle = \sum_\pi x_\pi(t) \otimes P(\pi) |{\bf v}\rangle
  \otimes |\uparrow_1 \ldots \uparrow_n\rangle = 2^{-n/2} \sum_\pi
  x_\pi(t) |{\bf v}\rangle \otimes |\pi\rangle,
\end{equation}
since $P(\pi)|\uparrow_1 \ldots \uparrow_n\rangle = 2^{-n/2}
|\pi\rangle$.  The reduced density matrix is
\begin{equation}\label{rra}
  \rho_r(t) = \mbox{Tr}_E [|\psi(t)\rangle\langle\psi(t)|] = 2^{-n}
  \sum_\pi x_\pi(t) P({\bf v}) x^\dagger_\pi(t).
\end{equation}
This can be further simplified by using the homomorphism between
$SU(2)$ and $SO(3)$. Define the rotation operators
\begin{equation}
  B_{+k}(t) = P({\bf u}_k) + \cos\theta_k(t) \overline{P}({\bf u}_k) -
  \sin\theta_k(t) {\bf u_k} \wedge,
\end{equation}
$B_{-k}(t) = B_{+k}^T(t)$ and $B_{\pi jk}(t) = B_{\pi_kk}(t) \ldots
B_{\pi_jj}(t)$. $B_{+k}(t)$ corresponds to a rotation of angle
$\theta_k(t)$ about ${\bf u}_k$, and $P({\bf u}_k) = {\bf u}_k {\bf
u}_k^T$, a projection operator on $R^3$. Note that $P({\bf u}_k)$ is
also used to indicate a projection in the system Hilbert space --- its
meaning should be clear from the context. $B_{\pi1n}(t)$ will usually
be simplified to $B_{\pi}(t)$. Then $x_{\pi_11}(t) P({\bf v})
x^\dagger_{\pi_11}(t) = P[B_{\pi_11}(t){\bf v}]$.  Eq.~(\ref{rra}) can
then be written
\begin{equation}\label{rrb}
  \rho_r(t) = 2^{-n} \sum_\pi P[B_\pi(t){\bf v}].
\end{equation}
Define $A_j(t) = 1/2[B_{+j}(t) + B_{-j}(t)] = P({\bf u}_j) +
\cos\theta_j(t) \overline P({\bf u}_j)$ and $A_{jk}(t) = A_k(t) \ldots
A_j(t)$, then $2^{-n}\sum_\pi B_\pi(t) = A_{1n}(t)$. $A_{1n}(t)$ will
usually be written $A(t)$.  Since $P[B_\pi(t){\bf v}]$ is linear in
$B_\pi(t)$ the sum in eq.~(\ref{rrb}) can then be done, so
\begin{equation}\label{rrc}
  \rho_r(t) = \frac{1 + \sigma.A(t){\bf v}}{2}.
\end{equation}
Generically this is not a projection operator since $|A(t){\bf v}|$
may not equal $1$. It is convenient however to define $P({\bf y}) =
1/2(1 + \sigma.{\bf y})$ for all ${\bf y} \in C^3$, and this extended
definition will be used throughout the paper. $P({\bf y})$ is a
projection operator if and only if ${\bf y}$ is a real unit vector.
Eq.~(\ref{rrc}) can now be written as $\rho_r(t) = P[A(t){\bf v}]$.

The eigenvalues of eq.~(\ref{rrc}) are $1/2[1\pm N(t)]$ and the
corresponding eigenstates, for $N(t) \neq 0$, are $|\pm {\bf
w}(t)\rangle$, where $N(t) = |A(t){\bf v}|$ and ${\bf w}(t) = A(t){\bf
v} N^{-1}(t)$.

Lemma $1$. \emph{Sufficient conditions that $N(t)\neq0$ for all $t$
  are that $\theta_i(t) \geq \theta_j(t)$ for all $i< j$ and ${\bf
  u}_i.  {\bf u}_{i+1} \neq 0$ for all $i\geq0$.}

Proof. Suppose $\exists t $ s.t.\ $N(t) = 0$, $\Rightarrow \mbox{det}
A(t) = 0$, $\Rightarrow \exists j $ s.t.\ $ \mbox{det} A_j(t) = 0$,
$\Rightarrow \theta_j(t) = \pi/2$. Let $j$ be the largest $j$ s.t.\
$\theta_j(t) = \pi/2$, then $A_i(t) = P({\bf u}_i) \forall i \leq j$
and $\mbox{det} A_i(t) \neq 0 \forall i>j$, $\Rightarrow N(t) = \|
A_{(j+1)n} (t) {\bf u}_j\| \prod_{j>i\geq0} |{\bf u}_i.{\bf u}_{i+1}|$
and $\mbox{det} A_{(j+1)n}(t) \neq 0$, $\Rightarrow \exists i$ s.t.\
$|{\bf u}_i.{\bf u}_{i+1}| = 0$ \#

For the rest of this paper it will be assumed that $\{\theta_i\}$ and
$\{{\bf u}_i\}$ satisfy the conditions of lemma $1$. The condition on
the $\{\theta_i\}$ holds so long as the environment spin particles are
further apart than the range of their individual interactions. The
condition on $\{{\bf u}_i\}$ holds generically and is physically
reasonable since any realistic experiment will not have exact
alignment.

\subsection{Decoherence matrix elements}
The Heisenberg picture Schmidt projection operators are
\begin{equation}
  P^\pm_H(t) = U^\dagger(t) P[{\bf \pm w}(t)] \otimes I_E U(t)
  \label{HSproja}.
\end{equation}
Eq.~(\ref{HSproja}) can be rewritten using eq.~(\ref{Uteq})
\begin{equation}
  P^\pm_H(t) = \sum_\pi x^\dagger_\pi(t) P[{\bf \pm w}(t)]
  x_\pi(t)\otimes P(\pi) = \sum_\pi P[{\bf \pm w}_\pi(t)] \otimes
  P(\pi), \label{HSprojb}
\end{equation}
where ${\bf w}_\pi(t) = B^T_\pi(t) {\bf w}(t)$.

Consider the probability of a history consisting of projections at
time $t$ and then $s$, where the projectors are Schmidt projectors.
\begin{equation}\label{proba}
  p(\pm\pm) = \| P^\pm_H(s) P^\pm_H(t) |\psi(0)\rangle\|^2.
\end{equation}
Eq.~(\ref{proba}) simplifies using eq.~(\ref{HSprojb}) and $ P(\pi)
|\psi(0)\rangle = 2^{-1/2} |{\bf v}\rangle \otimes |\pi\rangle$ to
become
\begin{eqnarray}
  p(\pm\pm) &=& \sum_\pi \| P[{\bf \pm w}_\pi(s)] P[{\bf \pm
    w}_\pi(t)] |{\bf v}\rangle \otimes P(\pi) |\uparrow_1 \ldots
    \uparrow_n\rangle\|^2 \nonumber \\ &=& 2^{-n-2} \sum_\pi [1 \pm
    {\bf w}_\pi(t).{\bf v}] [1 \pm {\bf w}_\pi(t). {\bf
    w}_\pi(s)]. \label{probb}
\end{eqnarray}
The off-diagonal decoherence matrix elements can be calculated
similarly.
\begin{eqnarray} 
  \lefteqn{\langle\psi(0)| P^\pm_H(t) P^\pm_H(s) P^\mp_H(t)
    |\psi(0)\rangle} && \nonumber \\ &=& 2^{-n} \sum_\pi \mbox{Tr} \{
    P({\bf v}) P[{\bf \pm w}_\pi(t)] P[{\bf \pm w}_\pi(s)] P[{\bf \mp
    w}_\pi(t)] \} \nonumber\\ &=& 2^{-n-2} \sum_\pi [{\bf w}_\pi (t)
    \wedge {\bf v}]. [ \pm {\bf w}_\pi (t) \wedge {\bf w}_\pi (s) \pm
    i {\bf w}_\pi (s)] \,. \label{twooffa}
\end{eqnarray}

For a general set of vectors $\{{\bf u}_k\}$ and time functions
$\{\theta_k\}$ eqs. (\ref{probb}) and (\ref{twooffa}) are very
complicated. However, with a restricted set of time functions a
complete analysis is possible. The functions $\{\theta_k\}$ are said
to describe a \emph{separated interaction} if, for all $t$, there
exists $k$ s.t. $\theta_j(t) = \pi/2$ for all $j<k$, and $\theta_j(t)
= 0$ for all $j>k$. For separated interactions a projection time $t$
is said to be \emph{between} interactions $j$ and $j+1$ when
$\theta_i(t) = \pi/2 $ for all $i \leq j$ and $\theta_i(t) = 0$ for
all $i > j$.  A projection time $t$ is said to be \emph{during}
interaction $j$ when $\theta_i(t) = \pi/2$ for all $i < j$,
$\theta_i(t) = 0$ for all $i > j$ and $0 < \theta_j(t) < \pi/2$.
Separated interactions have a simple physical meaning: the
interactions with the environment spins occur distinctly, and in
sequence.

Under this restriction a complete classification of all the consistent
sets, both branch dependent and branch independent, is possible. This
classification has a particularly simple form for generic ${\bf v}$
and $\{{\bf u}_k\}$ satisfying ${\bf u}_k.{\bf u}_{k+1} \neq 0$, and
${\bf u}_k \wedge {\bf u}_{k+1} \neq 0$ for all $k = 0, \ldots, n-1$.
Recall ${\bf u}_0 = {\bf v}$.  For weak consistency the second
requirement is stronger $({\bf u}_k \wedge {\bf u}_{k+1}).  ({\bf
u}_{k+2} \wedge {\bf u}_{k+1}) = {\bf u}_k \overline P ({\bf u}_{k+1})
{\bf u}_{k+1} \neq 0$. These assumptions will be assumed to hold
unless stated otherwise.

\subsection{Classification theorem}
The proof first considers projections at two times and shows that a
pair of times gives rise to non-trivial consistent histories only when
the earlier time is between interactions or the earlier time is during
an interaction and the later time between this interaction and the
next. The second part of the proof shows that any set of
branch-independent histories consisting of branches that satisfy this
rule for all pairs of projections is consistent. The proof holds for
weak and medium consistency criteria.

\subsubsection{Allowed histories}
Let $t$ be a time during interaction $j$. Define $\omega =
\theta_j(t)$ and $\phi = \theta_j(s)$. Define ${\bf x} = A_{1(j-1)}(s)
{\bf v} = A_{1(j-1)}(t) {\bf v}$ and ${\bf y} = A^T_{(j+1)n}(s)
A_{1n}(s) {\bf v}$.  Note $B_{\pi 1n}(t) = B_{\pi 1j}(t)$ and $B_{\pi
1(j-1)}(t) = B_{\pi 1(j-1)}(s)$ since $t<s$. With this notation and
using simple vector identities the off-diagonal elements of the
decoherence matrix (from eq.~\ref{twooffa}) are
\begin{equation} \label{twooffb}
  2^{-(n+2)} \sum_\pi [ {\bf w}(t) \wedge B_\pi(t){\bf v}].  [\pm {\bf
  w}(t) \wedge B_\pi(t) {\bf w}_\pi(s) \pm i B_\pi(t) {\bf w}_\pi(s)].
\end{equation}
Now
\begin{equation}\label{twooffba}
  B_\pi(t) {\bf w}_\pi (s) = B_{\pi j}(t) B_{\pi 1(j-1)}(t) B^T_{\pi
    1(j-1)}(s) B^T_{\pi jn}(s) {\bf w}(s) = B_{\pi j}(t) B^T_{\pi
    jn}(s) {\bf w}(s),
\end{equation}
which only depends on $\pi_i$ for $i \geq j$. Since $B_{\pi 1j}(t){\bf
  v}$ only depends on $\pi_i$ for $i \leq j$ the sum
  eq.~(\ref{twooffba}) can be done over all $\pi_i$, $i \neq j$.
\begin{eqnarray}
  2^{1-j} \sum_{\pi_i,\, i<j} B_{\pi 1j}(t){\bf v} &=& [A_j (t) -
  \pi_j \sin\omega\, {\bf u}_j \wedge] A_{1(j-1)}(t) {\bf v}\\ &=&
  {\bf w}(t) N(t) - \pi_j \sin\omega\, {\bf u}_j \wedge {\bf x},
\end{eqnarray}
\begin{eqnarray}
  2^{-(n-j)} \sum_{\pi_i,\, i>j} B_{\pi j}(t) B^T_{\pi jn}(s){\bf
  w}(s) &=& N^{-1}(s) B_{\pi j}(t) B_{\pi j}^T(s) A^T_{(j+1)n}(s)
  A_{1n}(s){\bf v} \\ &=& N^{-1}(s) B_{\pi j}(t) B_{\pi j}^T(s) {\bf
  y}.
\end{eqnarray}
Substitute these last two results into eq.~(\ref{twooffb}) which
becomes
\begin{eqnarray}\nonumber
  \lefteqn{2^{-3} N^{-1}(s) \sum_{\pi_j} \{{\bf w}(t) \wedge [{\bf
    w}(t) N(t) - \pi_j \sin\omega {\bf u}_j \wedge {\bf x}]\}}
    \hspace{2in} \\ &&.  \label{twooffc} [\pm {\bf w}(t) \wedge B_{\pi
    j}(t) B_{\pi j}^T(s){\bf y} \pm i B_{\pi j}(t) B_{\pi j}^T(s){\bf
    y}].
\end{eqnarray}
This can easily be simplified since ${\bf w}(t) \wedge {\bf w}(t) =
0$. The only remaining term in the first bracket is then linear in
$\pi_j$, so when the sum over $\pi_j$ is taken only the terms linear
in $\pi_j$ in the second bracket remain. Eq.~(\ref{twooffc}) is
therefore
\begin{equation} \label{twooffd}
  1/4 N^{-1}(s) \sin\omega \sin(\omega-\phi) [{\bf w}(t) \wedge ({\bf
    u}_j \wedge {\bf x})]. [{\bf w}(t) \wedge ({\bf u}_j \wedge {\bf
    y}) \pm i {\bf u}_j \wedge {\bf y}].
\end{equation}
Now ${\bf w}(t) = [P({\bf u}_j) + \cos\omega \overline P({\bf u}_j)]
{\bf x} N^{-1}(t)$ so ${\bf w} (t). ({\bf x} \wedge {\bf u}_j) =
0$. Therefore
\begin{equation}\label{s1}
  [{\bf w}(t) \wedge ({\bf u}_j \wedge {\bf x})]. [{\bf w}(t) \wedge
  ({\bf u}_j \wedge {\bf y})] ={\bf x}^T \overline P ({\bf u}_j) {\bf
  y}.
\end{equation}
Also ${\bf u}_j. {\bf w}(t) = {\bf u}_j. {\bf x}N^{-1}(t)$ so
\begin{equation}\label{s2}
  [{\bf w}(t) \wedge ({\bf u}_j \wedge {\bf x})]. ({\bf u}_j \wedge
  {\bf y}) = - N^{-1}(t) ({\bf u}_j. {\bf x}) {\bf x}. ({\bf u}_j
  \wedge {\bf y}).
\end{equation}
Eq.~(\ref{twooffc}) can be simplified using eq.~(\ref{s1}) and
eq.~(\ref{s2}) to
\begin{equation} \label{twooffe}
  1/4 N^{-1}(s) \sin\omega \sin(\phi-\omega) \{ \pm {\bf x}^T
  \overline P ({\bf u}_j) {\bf y} \pm i N^{-1}(t) ({\bf u}_j. {\bf x})
  {\bf u}_j. ({\bf x} \wedge {\bf y}) \}
\end{equation}
The probabilities can be calculated during the same results. Summing
all the terms $i \neq j$ in eq.~(\ref{probb}) results in
\begin{eqnarray}\nonumber
  &&2^{-3} \sum_{\pi_j} \{1 \pm {\bf w}(t).[{\bf w}(t) N(t) - \pi_j
  \sin\omega {\bf u}_j \wedge {\bf x}]\} \left\{ 1 \pm \frac{{\bf x}^T
  A_j(\omega) B_{\pi j}(t) B^T_{\pi j}(s) {\bf y}}{N(s)N(t)}\right\}
  \\ &=& 2^{-2} [1 \pm N(t)] \left\{1 \pm \frac{{\bf x}^T [P({\bf
  u}_j) + \cos\omega\cos(\phi-\omega) \overline P({\bf u}_j)] {\bf
  y}}{N(s)N(t)}\right\}\label{probc}
\end{eqnarray}
$N^2(s) = |A_{1n}(s){\bf v}| = {\bf x}^T A_j(\phi) {\bf y}$ and
  $\cos(\omega-\phi) \cos\omega - \cos\phi =
  \sin\omega\sin(\phi-\omega)$, so eq.~(\ref{probc}) is
\begin{equation}\label{probd}
  1/4[1 \pm N(t)] \left[1 \pm \frac{N^2(s) + \sin\omega
    \sin(\phi-\omega){\bf x}^T \overline P({\bf u}_j) {\bf y}}
    {N(s)N(t)}\right]
\end{equation}

To write the decoherence matrix without using ${\bf x}$ and ${\bf y}$
it is necessary to consider three cases: when times $s$ and $t$ are
during the same interaction, when they are during adjacent
interactions and when they are during separated interactions. If $t$
is during interaction $j$ and $s$ during interaction $k$ the three
cases are $k=j$, $k=j+1$ and $k>j+1$. For the remainder of this
section let $\phi = \theta_k(s)$,
\begin{equation}
  N_j(\omega) = |A_j(t) {\bf u}_{j-1}| \mbox{~and~} \lambda_{ij} =
  \prod_{j>k\geq i} |{\bf u}_k. {\bf u}_{k+1}|\,.
\end{equation}
Then
\begin{eqnarray}
  {\bf x} &=& \lambda_{0(j-1)} {\bf u}_{j-1} \\ N(t) &=&
  \lambda_{0(j-1)} N_j(\omega) \\ N(s) &=& \lambda_{0(k-1)} N_k(\phi)
  \\ {\bf y} &=& \left\{
  \begin{array}[c]{ll}
    \lambda_{0(j-1)} A_j(s) {\bf u}_{j-1} & \mbox{for $k=j$} \\
    \lambda_{0j} A_{j+1}^2(s) {\bf u}_{j} & \mbox{for $k=j+1$} \\
    \lambda_{(j+1)(k-1)} \lambda_{0(k-1)} N^2_k(\phi) {\bf u}_{j+1} &
    \mbox{for $k>j+1$}
  \end{array} \right.
\end{eqnarray}
The probabilities of the histories (eq.~\ref{probc}) are
\begin{eqnarray}
  p(\pm\pm) = 1/4 [1 \pm \lambda_{0(j-1)} N_j(\omega)] [1 \pm a]
\end{eqnarray}
where
\begin{eqnarray}
  a = \left \{
    \begin{array}[c]{ll}
      \frac{N_j^2(\phi) + \sin\omega \cos\phi \sin(\phi-\omega) |{\bf
          u}_{j-1} \wedge {\bf u}_j|^2}{ N_j(\omega) N_j(\phi)} &
          \mbox{for $k=j$} \\ \frac{ \lambda_{(j-1)j} N_{j+1}^2(\phi)
          + \cos\omega \sin\omega \lambda^2_{j(j+1)} \sin^2\phi {\bf
          u}_{j-1}^T \overline P({\bf u}_j) {\bf u}_{j+1}}{
          N_j(\omega) N_{j+1}(\phi)} &\mbox{for $k=j+1$} \\
          N_{k}(\phi)\frac{ \lambda_{(j-1)(k-1)} +
          \lambda_{(j+1)(k-1)} \cos\omega \sin\omega {\bf u}_{j-1}^T
          \overline P({\bf u}_j) {\bf u}_{j+1}}{ N_j(\omega) }
          &\mbox{for $k>j+1$}
    \end{array} \right..
\end{eqnarray}
The nonzero off-diagonal terms are (eq.~\ref{twooffe})
\begin{equation} \label{twoofff}
\left \{
  \begin{array}[c]{ll}
    \frac{\lambda_{0(j-1)} \sin\omega \sin(\phi-\omega) \cos\phi |{\bf
        u}_{j-1} \wedge {\bf u}_j|^2}{4 N_j(\phi)} & \mbox{for
        $k=j$}\\ \frac{\lambda_{0(j-1)} \lambda_{j(j+1)} \sin\omega
        \cos \omega \sin^2 \phi[ N_j(\omega) {\bf u}_{j-1}^T \overline
        P({\bf u}_j) {\bf u}_{j+1} \pm i \lambda_{(j-1)j} {\bf
        u}_{j-1}.({\bf u}_j \wedge {\bf u}_{j+1})]}{4 N_j(\omega)
        N_{j+1}(\phi)} & \mbox{for $k=j+1$}\\ \frac{\lambda_{0(j-1)}
        \lambda_{(j+1)(k-1)} N_k(\phi) \sin\omega \cos \omega[
        N_j(\omega) {\bf u}_{j-1}^T \overline P({\bf u}_j) {\bf
        u}_{j+1} \pm i \lambda_{(j-1)j} {\bf u}_{j-1}.({\bf u}_j
        \wedge {\bf u}_{j+1})]}{4 N_j(\omega)} & \mbox{for $k>j+1$.}
  \end{array} \right.
\end{equation}

The off-diagonal terms can be zero for two reasons, either there is a
degeneracy in the measurement spin directions, or $s$ and $t$ take
special values. The necessary and sufficient conditions for the
measurement spin directions not to be degenerate is that for all $j$
${\bf u}_j. {\bf u}_{j+1} \neq 0$ and ${\bf u}_j \wedge {\bf u}_{j+1}
\neq 0$. The first condition ensures that $\lambda_{ij} \neq 0$ for
all $i$ and $j$ and that the Schmidt states are well defined.  These
cases do not need to be considered when we are interested in exact
consistency because they have measure zero and \emph{almost surely}
under any perturbation the degeneracy will be lifted. If weak
consistency is used only the real part needs to vanish and the
measurement direction need to satisfy the stronger condition ${\bf
u}_{j-1}^T \overline P({\bf u}_j) {\bf u}_{j+1} \neq 0 $ for all
$j$. This is still of measure zero. If approximate consistency is
being considered the situation is more complicated as the histories
will remain approximately consistent under small enough perturbations.
This will not be considered in this letter.  Unless said otherwise it
will be assumed that the measurement spin direction are not
degenerate.

Therefore from eqs.~(\ref{twoofff}) the only pairs of times giving
rise to consistent projections are repeated projections (that is $s=t$
which implies $j=k$ and $\omega=\phi$), projections in between
interactions and any later time (that is $\omega = 0$ or $\pi/2$), and
a projection during an interaction and a projection at the end of the
same interaction (that is $j=k$ $\omega \in [0,\pi/2]$ and $\phi=
\pi/2$.)

\subsubsection{Probabilities of allowed histories}
The model is invariant under strictly monotonic reparameterisations of
time, $t \to f(t)$. Therefore for separated interactions no generality
is lost by choosing the time functions $\{\theta_j\}$ such that the
$j^{\mbox{\scriptsize th}}$ interaction finishes at $t=j$, that is
$\theta_i(j) = \pi/2$ for all $i \leq j$ and $\theta_i(j) = 0$ for all
$i>j$. It is convenient to define $R_{\pi ij} = [P({\bf u}_i) - \pi_i
{\bf u}_i \wedge] \ldots [P({\bf u}_i) - \pi_i {\bf u}_i \wedge]$.
Then $B_\pi(m) = R_{\pi 1m}$.

Consider the history $\alpha$ that consists of projections at times
$\{m_i: i = 1,2, \ldots l\}$, then at time $t \in (k-1,k)$ and then at
time $k$, where $\{m_i,k\}$ is an ordered set of positive integers.
This history means that the particle spin was in direction $\pm{\bf
u}_{m_i}$ at time $m_i$, $i = 1, \ldots ,l$, direction $\pm{\bf w}(t)$
at time $t$ and direction $\pm{\bf u}_k$ at time $k$.  Define ${\bf
u}_0 = {\bf v}$ and $m_0 = 0$.

Using the same method as for two projections the probability for
history $\alpha$ is
\begin{eqnarray} \nonumber
  p_\alpha &=& 2^{-n} 2^{-(l+2)} \sum_\pi \prod_{i=0}^{l-1} [1 +
  \alpha_{i} \alpha_{i+1} {\bf w}_\pi(m_{i}).  {\bf w}_\pi(m_{i+1})]
  \\ && \mbox{} \times [1 + \alpha_l \alpha_{t} {\bf w}_\pi(m_{l}).
  {\bf w}_\pi(t)] \times [1 + \alpha_{t} \alpha_{k} {\bf w}_\pi(t).
  {\bf w}_\pi(m_k)]
  \label{probsa}
\end{eqnarray}
Now
\begin{eqnarray}
  {\bf w}_\pi(m_{i}). {\bf w}_\pi(m_{i+1}) = {\bf u}_{m_{i}}^T R_{\pi
  1 m_i} R^T_{\pi 1 m_{i+1}} {\bf u}_{m_{i+1}} = {\bf u}_{m_{i}}^T
  R_{\pi (m_{i}+1) m_{i+1}} {\bf u}_{m_{i+1}},
\end{eqnarray}
which only depends on $\pi_j$ for $m_{i+1} \geq j > m_i$.  Also
\begin{equation}
  {\bf w}_\pi(t).{\bf w}_\pi(k) = N^{-1}_k(t) {\bf u}_{k-1}^T A_k(t)
  B_{k\pi_k}(t) {\bf u}_{k} = N^{-1}_k(t) ({\bf u}_{k-1}. {\bf
  u}_{k}),
\end{equation}
which is independent of $\pi$ and
\begin{equation}
  {\bf w}_\pi(t).{\bf w}_\pi(m_l) = N^{-1}_k(t) {\bf u}_{k-1}^T A_k(t)
  B_{\pi_k k}(t) R_{\pi (m_l+1) (k-1)} {\bf u}_{m_l},
\end{equation}
which only depends on $\pi_j$ for $j > m_l$.  These last three
equations show that each $B_{\pi_i i}$ is linear so the sum over $\pi$
is trivial and each $B_{\pi_i i}$ can be replaced by $A_i$.
\begin{equation}
  2^{m_{i}-m_{i+1}-1}
  \sum_{\makebox[0in][c]{\scriptsize$\pi_j,\,m_{i+1}>j>m_i$}} {\bf
  w}_\pi(m_{i}). {\bf w}_\pi(m_{i+1}) = {\bf u}_{m_{i}}^T P({\bf
  u}_{m_{i}+1}) \cdots P({\bf u}_{m_{i+1}-1}) {\bf u}_{m_{i+1}} =
  \lambda_{m_{i}m_{i+1}},
\end{equation}
\begin{equation}
  2^{m_l-k} \sum_{\makebox[0in][c]{\scriptsize$\pi_i,\,k \geq i >
  m_l$}} {\bf w}_\pi(t). {\bf w}_\pi(m_l) = N^{-1}_k(t) {\bf
  u}_{k-1}^T A^2_k(t) {\bf u}_{k-1} \lambda_{m_l(k-1)} =
  \lambda_{m_l(k-1)} N_k(t)
\end{equation}
Using these results to do the sum over all $\pi$ eq.~(\ref{probsa}) is
\begin{equation}\label{probsb}
  p_\alpha = 2^{-(l+2)} [1 + \alpha_l \alpha_{t} \lambda_{m_l(k-1)}
  N_k(t)] [1 + \alpha_t \alpha_{k} N^{-1}_k(t) ({\bf u}_{k-1}. {\bf
  u}_k)] \prod_{i=0}^{l-1} [1 + \alpha_i \alpha_{i+1}
  \lambda_{m_im_{i+1}}].
\end{equation}
 
\subsubsection{Consistency of allowed histories}
Since a coarse graining of a consistent set is consistent it is
sufficient to only consider the off-diagonal decoherence matrix
elements between the most finely grained allowed histories, which are
those that consist of projections between all interactions and one
projection during the interaction before the final projection. The
off-diagonal elements of the decoherence matrix arise from only three
forms, which depend on where the two branches separate, that is the
earliest projector where they differ.

First consider the case where two histories differ at a projection in
between interactions and all projections up to that point have also
been in between interactions. Let $C_\alpha = Q_\alpha P_H(k) \ldots
P_H(1)$ and $C_\beta = Q_\beta \overline P_H(k) \ldots P_H(1)$. The
decoherence matrix element between them is
\begin{eqnarray}\nonumber
  2^{-n} \sum_\pi \mbox{Tr} \{Q_\pi P({\bf u}_k) x_\pi(k) P[{\bf
    w}_\pi(k-1)] \ldots P[{\bf w}_\pi(1)] P({\bf v}) \\ \times P[{\bf
    w}_\pi(1)] \ldots P[{\bf w}_\pi(k-1)] x^\dagger_\pi(k) P(-{\bf
    u}_k)\} \label{offcase1a}
\end{eqnarray}
where $Q_\pi = \langle\pi| x_\pi(k)Q^\dagger_\alpha Q_\beta
x^\dagger_\pi(k) |\pi\rangle$. Since $Q_\alpha$ and $Q_\beta$ only
contain projections after interaction $k$ has completed $Q_\pi$ is
independent of $\pi_j$ for all $j \leq k$.  Now $P[{\bf w}_\pi(j)]
P[{\bf w}_\pi(j-1)] P[{\bf w}_\pi(j)] = 1/2(1+ {\bf u}_{j-1}. {\bf
u}_j) P[{\bf w}_\pi(j)]$. Let $\mu = 2^{1-m}\prod_{0<j<m} (1+ {\bf
u}_{j-1}.{\bf u}_j)$ and eq.~(\ref{offcase1a}) is
\begin{equation} \label{offcase1b}
  \mu 2^{-n} \sum_\pi \mbox{Tr} \{ Q_\pi P({\bf u}_k) P[B_\pi(k) {\bf
    w}_\pi(k-1)] P({\bf -u}_k)\}
\end{equation}
But $1/2\sum_{\pi_k} P[B_\pi(k) {\bf w}_\pi(k-1)] = P[{\bf u}_k({\bf
  u}_k.{\bf u}_{k-1})]$ and $P({\bf u}_k) P[{\bf u}_k({\bf u}_k.{\bf
  u}_{k-1})] P({\bf -u}_k) = 0$ so eq.~(\ref{offcase1b}) is zero.

Now consider $C_\alpha = P_H(k) P_H(t) P_H(k-1) \ldots P_H(1)$ and
$C_\beta = P_H(k) \overline P_H(t) P_H(k-1) \ldots P_H(1)$. The
decoherence matrix element between them is
\begin{equation} \label{offcase2a}
  \mu 2^{-n} \sum_\pi \mbox{Tr} \{P[{\bf w}_\pi(k)] P[{\bf w}_\pi(t)]
  P[{\bf w}_\pi(k-1)] P[{\bf -w}_\pi(t)] P[{\bf w}_\pi(k)]\},
\end{equation}
which, because $B_{\pi_kk} {\bf u}_k = {\bf u}_k$ equals
\begin{equation}
  \mu 2^{-n} \sum_\pi \mbox{Tr} \{P({\bf u}_k) P[{\bf w}(t)]
  P[B_{\pi_kk}(t) {\bf u}_{k-1}] P[{\bf -w}(t)] P({\bf w}(k)\}.
  \label{offcase2b}
\end{equation}
The sum over $\pi_k$ can be done to give $P[{\bf w}(t)] P[A_k(t) {\bf
  u}_{k-1}] P[{\bf -w}(t)]$, and since ${\bf w}(t)$ is parallel to
  $A_k (t) {\bf u}_{k-1}$, eq.~(\ref{offcase2b}) is zero.

The final case to consider is when then the histories $\alpha$ and
$\beta$ differ in their final projection. They will be trivially
consistent.

\section{The algorithm applied to the spin model}\label{sec:spininf}

A set of histories that maximises information must be complete,
therefore all histories must consist of projections at times
$\{1,\ldots, k-1 ,t ,k: t \in(k-1,k)\}$. First we show that $k$ must
be the same for all histories, then we show that generically $k=n$.
That is, the algorithm selects a branch independent set that
generically describes a measurement at the end of each interaction
plus one measurement during the final interaction.

The information content of two subtrees rooted at the same point only
depends on the projection times within each one. Either the two
subtrees have the same information, in which case their projection
times must be the same, or one has more, but since the projection
times used in the subtree with greater information will also be
consistent if used in the subtree with less information these
projection times can be used instead. Therefore in the set with
maximum information all the subtree must have the same projection
times, thus all the histories must have the same projection times ---
the maximal set is branch independent.

Let the projection times be $\{1,\ldots, k-1 ,t ,k: t \in(k-1,k)\}$.
Then from eq.~(\ref{probsb}) and eq.~(\ref{infadd}) the information
content of this set is

\begin{eqnarray} \label{kinfa}
  f[N_k(\theta_k(t))] + f[({\bf u}_{k}.  {\bf u}_{k-1}) N_k^{-1}
    (\theta_k(t))] + \sum_{k>j>0} f({\bf u}_{j-1}. {\bf u}_j)
\end{eqnarray}
where
\begin{equation}
  f(x) = - \frac{1+x}{2} \log \frac{1+x}{2} - \frac{1-x}{2} \log
  \frac{1-x}{2}.
\end{equation}
Maximising eq.~(\ref{kinfa}) with respect to $t$ yields
\begin{equation}
  E({\cal S}_k) = E_k = 2f(|{\bf u}_{k}.  {\bf u}_{k-1}|^{1/2}) +
  \sum_{k>j>0} f({\bf u}_{j-1}. {\bf u}_j),
\end{equation}
where ${\cal S}_k$ is the branch independent set consisting of
projections at times $\{1,\ldots,k-1,t_k,k\}$.  This is usually
maximised by $k=n$ but depending on the relationships between the
${\bf u}_j$ any value of $k$ may be possible. For example, consider
${\bf u}_{j-1}.{\bf u}_j = 1-\epsilon$ for all $j \neq k$ and ${\bf
u}_{k-1}. {\bf u}_k = \epsilon$ and $\epsilon$ is small.
\begin{equation}
  E_m = \left\{
    \begin{array}{lr}
      O(\epsilon\log \epsilon), & \mbox{for $m < k$}, \\ 2 \log 2 +
      O(\epsilon\log \epsilon), & \mbox{for $m = k$}, \\ \log 2 +
      O(\epsilon\log \epsilon) & \mbox{for $m > k$},
    \end{array}\right.
\end{equation}
which for small $\epsilon$ is maximised by $E_k$.

The precise relationship between the $\{{\bf u}_j\}$ that ensure $E_n
E_k$ for all $k < n$ is complicated in detail, but simple
qualitatively. Roughly speaking, $E_n < E_k$ only if $|{\bf u}_{j-1}.
{\bf u}_j| \gg |{\bf u}_{k-1}.  {\bf u}_k|$ for all $j > k$, that is
all the measurement directions must be approximately parallel after
the $k^{\mbox{\scriptsize th}}$.  Monte Carlo integration over $\{{\bf
u}_i\}$ (with the $SO(3)$ invariant measure) shows that for $n=3$ set
${\cal S}_n$ is selected $85.7\%$ of the time, for $n=4$ it is
selected $84.3\%$ of the time, and for all $n>4$ it is selected
$84.2\%$ of the time.  When the vectors are approximately parallel,
that is $|{\bf u}_{j-1}.  {\bf u}_j| = 1 - O(\epsilon)$, set $S_n$ is
selected with probability $1-O(\epsilon)$.  If however all the
measurement spins are approximately parallel ($|{\bf u}_{j-1}.  {\bf
u}_j| > 1 - \epsilon$, and $-n\epsilon\log\epsilon<4\log2$) then for
some orientations of the initial system spin (${\bf v} = {\bf u}_0$)
$E_1 > E_k$ for all $E_k$ so set ${\cal S}_1$ is selected.  That is,
the maximal set consists only of a projection during the first
interaction and at the end of the first interaction.

Though the results of the algorithm may seem counterintuitive the
following discussion shows why this is not a problem.

First consider the case when the system is genuinely closed. All the
projections before the last interaction are natural, in the sense that
they agree with our intuitive understanding of a measurement type
process. It is only the projections during the last interaction, which
occur when the set of histories is nearly complete, that are
unnatural. Our intuition about the system and the result we believe to
be correct relies on the experiment being embedded in a larger system
in which the sets of histories considered are always far from
complete.

Second consider the case where the system is approximately closed.
Then the sets ${\cal S}_k$ should describe the first projections of a
maximum-information solution in a larger Hilbert space. For reasons
explained below, no non-trivial projections onto the system space will
result in consistent extensions of the sets ${\cal S}_k$, even if the
system interacts with new degrees of freedom in the environment.  This
shows that though it is a maximum-information set for a subsystem, it
is unlikely to be part of the maximum-information set for the entire
system. The set most likely to be part of the maximum-information set
is the natural set, the set that consists of projections only at the
end of each interaction.

The set of normalised histories (in the Schr\"odinger picture at time
$k$, that is the path-projected states) is
\begin{equation}
    {\cal S}_k = \{ |\alpha_0{\bf v}_k \rangle \otimes
    |\alpha_1(\uparrow), \ldots, \alpha_{k-1}(\uparrow),
    \alpha_k(\rightarrow), \uparrow_{k+1}, \ldots, \uparrow_n\rangle
    \forall \alpha \in Z_2^{k+1}\},
\end{equation}
where $\alpha$ is a string of $2^{k+1}$ plusses and minuses,
$+(\uparrow) = \uparrow$, $-(\uparrow) = |\downarrow\rangle$ and
$\pm(\rightarrow)$ are orthogonal vectors depending on ${\bf u}_{k-1}$
and ${\bf u}_k$. This set of histories cannot be non-trivially
extended with Schmidt projections (see sec.~\ref{sec:spin:analysis}).
The reason for this is clear.  Consider two of the histories $|{\bf
\pm v}_k \rangle \otimes |e\rangle$ where $|e\rangle$ is the
environment state. These histories are only orthogonal because of the
orthogonality of the system part of the states. There can be no future
non-trivial extensions unless there is an exact degeneracy, because
consistency terms between these two histories will contain terms like
$|\langle{\bf v}| P({\bf w}) |{\bf v}\rangle| = \sqrt{1/2(1+{\bf
v}.{\bf w})}$, which is only zero when ${\bf w}=-{\bf v}$. In contrast
if projections are only made at the end of interactions all the
histories are orthogonal in the environment Hilbert space of the
finished interactions. Unless these interactions are ``undone'' these
histories will always remain orthogonal and cannot interfere. This
argument suggests that the true maximum-information set for the total
Hilbert space starts of with projections at the end of every
interaction but at \emph{no} interior times.

This suggests that an algorithm designed to produce a
maximum-information set for a subsystem could be constructed by
requiring that all the histories in a set were orthogonal in the
environment space, that is the reduced density matrices in the
environment Hilbert space for each history are orthogonal. This is
equivalent to considering sets of histories that satisfy the strong
consistency criterion (\ref{GMH:strong}) when the set $\{{\cal
M}_\alpha\}$ is chosen to be $\{P \otimes I: \mbox{for all projectors
$P$ on ${\cal H}_1$}\}$.

\section{Other algorithms}\label{sec:otheralg}

Let ${\cal G}({\cal H},U,|\psi\rangle)$ be the class of all sets of
non-trivial, exactly consistent, branch-dependent histories formed
{}from Schmidt projection operators in the spin model.  Consider an
algorithm that selects the set in ${\cal G}$ that minimises Isham and
Linden's information-entropy~(\ref{IL:information}). Due to the
special symmetries of the spin model the selected set will be branch
independent --- the argument at the start of
section~(\ref{sec:spininf}) is valid.

Consider the set of projections at $m$ times, so that the normalised
dimension of each history is $2^{-m}$. Information-entropy for this
set is
\begin{equation}
  S' = - \sum_{\alpha \in {\cal S}} p_{\alpha} \log
  \frac{p_\alpha}{(1/2)^{2m}} = -2m \log (2) - \sum_{\alpha \in {\cal
  S}} p_{\alpha} \log p_\alpha\,.
\end{equation}
Using the notation of the previous section this can be written
\begin{equation}
  S'' = -\sum_{m > k > 0} [2 \log 2 - f(\alpha_k)]\,,
\end{equation}
where the $\alpha_k$ depend on the projection times and vary between
$-1$ and $1$.  Since $f(x)\leq \log 2$ each term in the sum is always
negative so the minimum occurs for $m = n+1$, and the selected set
consists of projections at the end of every interaction and a
projection either at the end or the beginning of the last interaction
--- the algorithm has selected a natural set. The
minimum-information-entropy algorithm selects a set with as many
projections as possible, and among these sets it selects the set whose
probabilities have the lowest Shannon information. One drawback with
this approach is that unless trivial histories are excluded, or the
number of histories in a set bounded, the minimum may not exist and
the algorithm would therefore be ill defined. In particular if an
infinite number of repeated projections are allowed the algorithm is
ill defined.

\section{Conclusions}

This paper defines a precise algorithm for making probabilistic
predictions for closed quantum systems. The algorithm considers the
class of all non-trivial, exactly consistent, branch-dependent sets of
histories defined by Schmidt projections with respect to a fixed split
of the Hilbert space and selects from among them the set with the
maximum Shannon information. The algorithm avoids many of the problems
of the algorithms considered in ref.~\cite{McElwaine:Kent}. Because it
considers the entire time evolution of a system -- roughly speaking it
is global in time, whereas the algorithms in
ref.~\cite{McElwaine:Kent} are local --- it does not make unphysical
projections in systems where recoherence occurs and it produces
complete sets of histories that describe the correlations between the
system and the environment. Trivial and very small probability
histories, which cause problems for some of the algorithms considered
in ref.~\cite{McElwaine:Kent} by preventing later physical
projections, are unlikely to be selected since they contribute little
information. The algorithm is also likely to be stable under
perturbations in the initial conditions, the Hamiltonian and the
parameters, since it involves maximising a continuous function.

Section~\ref{sec:spininf} has shown that the algorithm selects a
natural set for a simple spin model.  It would be interesting to test
out the algorithm on more realistic examples; however, it seems
difficult to apply the algorithm directly, because of the large size
and complicated nature of ${\cal G}$.  Analytic calculations are only
possible when the system is very simple and in more realistic examples
computer simulations will be necessary.  However, it should be
possible at least to get some insight into the algorithm's predictions
by maximising subject to constraints, that is by considering a more
computationally tractable subset of ${\cal G}$.  For example, we could
choose a time interval $T$ that is greater than the time of individual
interactions (within the particular system) and larger than any
timescale over which recoherence occurs.  This would be used as a
moving time-window over which to perform the maximisation. The
earliest projection within each time-window would be selected and the
next time-window would commence from that time. Such algorithms should
select the same set as a global algorithm if $T$ is large enough, and
are also independently interesting.

Because the algorithm predicts the probabilities for events \emph{and}
the set of possible events the algorithm is falsifiable: the algorithm
is wrong if it selects any sets that do not agree with our
experiences. The algorithm can also be applied to situations where we
have no experience of what the natural sets of histories are: for
example, a (finite) closed system of electrons and photons --- and
perhaps ultimately could be applied to theories of quantum cosmology.


\begin{thebibliography}{10}

\bibitem{Dowker:Kent:properties}
F. Dowker and A. Kent, Phys. Rev. Lett. {\bf 75},  3038  (1995).

\bibitem{Dowker:Kent:approach}
F. Dowker and A. Kent, J. Stat. Phys. {\bf 82},  1575  (1996).

\bibitem{McElwaine:Kent}
A. Kent and J.~N. McElwaine, Quantum Prediction Algorithms, 
gr-qc/9610028,
  DAMTP/96-88, submitted to Phys. Rev. A.

\bibitem{McElwaine:2}
J.~N. McElwaine, Ph.D. thesis, DAMTP, Cambridge University, 1996.

\bibitem{gmhstrong}
M. Gell-Mann and J.~B. Hartle, gr-qc/9509054, 
{U}niversity of California, Santa
  Barbara preprint UCSBTH-95-28.

\bibitem{Isham:Linden:information}
C.~J. Isham and N. Linden, Information-entropy and the space of 
decoherence
  functions in generalised quantum theory, {I}mperial/TP/95-96/63,
  DAMTP/R96-44, submitted to Phys. Rev. A and to appear in quant-ph.

\bibitem{omnesbook}
R. Omn\`es, {\em The Interpretation of Quantum Mechanics} 
(Princeton University
  Press, Princeton, 1994).

\bibitem{griffithschqr}
R.~B. Griffiths, quant-ph/9606004, to appear in Phys. Rev. A.

\bibitem{CEPI:GMH}
M. Gell-Mann and J.~B. Hartle,  in {\em Complexity, 
Entropy and the Physics of
  Information}, Vol.~III of {\em SFI Studies in the 
Science of Complexity},
  edited by W.~H. Zurek (Addison Wesley, Reading, 1990).

\bibitem{Kent:implications}
A. Kent, gr-qc/9607073, {D}AMTP/96-74, submitted to Ann. Phys.

\bibitem{Goldstein:Page}
S. Goldstein and D.~N. Page, Phys. Rev. Lett. {\bf 74},  
3715  (1995).

\bibitem{Dowker:Halliwell}
H.~F. Dowker and J.~J. Halliwell, Phys. Rev. D {\bf 46},  
1580  (1992).

\bibitem{McElwaine:1}
J.~N. McElwaine, Phys. Rev. A {\bf 53},  2021  (1996).

\bibitem{Albrecht:decoherence}
A. Albrecht, Phys. Rev. D {\bf 46},  5504  (1992).

\bibitem{Albrecht:collapsing}
A. Albrecht, Phys. Rev. D {\bf 48},  3768  (1993).

\bibitem{McElwaine:4}
J.~N. McElwaine, Chapter 6 in \cite{McElwaine:2}.

\bibitem{Hartle:spacetime:information}
J.~B. Hartle, Phys. Rev. D {\bf 51},  1800  (1995).

\bibitem{Shannon:Weaver}
C.~E. Shannon and W. Weaver, {\em The Mathematical 
Theory of Communication}
  (University of Illinois, Urbana, 1949), 5th Edition, 1972.

\bibitem{Kent:information}
A. Kent, gr-qc/9610075, {D}AMTP/96-93, submitted to Phys. Rev. D.

\bibitem{Jaynes:papers}
E.~T. Jaynes,  in {\em Papers on Probability, Statistics and 
Statistical
  Mechanics}, edited by R.~D. Rosenkrantz (Reidel, Dordrecht, 1983).

\bibitem{Isham:homog}
C.~J. Isham, J. Math. Phys. {\bf 35},  2157  (1996).

\bibitem{Isham:Linden:temporal}
C.~J. Isham and N. Linden, J. Math. Phys. {\bf 35},  6360  (1994).

\bibitem{Isham:Linden:Schreckenberg}
C.~J. Isham, N. Linden, and S. Schreckenberg, 
J. Math. Phys. {\bf 35},  6360
  (1994).

\bibitem{Zurek:algorithmic}
W.~H. Zurek, Phys. Rev. A {\bf 40},  4731  (1989).

\end{thebibliography}
\end{document}